\documentclass[prd, superscriptaddress, tightenlines, longbibliography, nofootinbib, eqsecnum, amsfonts, amsmath, floatfix, twocolumn, notitlepage]{aa}

\usepackage[utf8]{inputenc}
\usepackage{mathrsfs}
\usepackage{euscript}
\usepackage{graphics}
\usepackage{graphicx}
\usepackage{amsmath}
\usepackage{amssymb}
\usepackage{gensymb}
\usepackage{bm}
\usepackage[usenames,dvipsnames,svgnames,table]{xcolor}
\definecolor{linkcolor}{rgb}{0.6,0,0}
\definecolor{citecolor}{rgb}{0,0,0.75}
\definecolor{urlcolor}{rgb}{0.12,0.46,0.7}
\usepackage{xspace}
\usepackage{wasysym}
\usepackage{times}
\usepackage{appendix}
\usepackage{comment}
\usepackage{lipsum}
\usepackage[nolist,nohyperlinks]{acronym}
\usepackage{float}
\usepackage{simplewick}
\usepackage[varg]{txfonts}
\usepackage{natbib, ifthen}
\usepackage[breaklinks, colorlinks, urlcolor=urlcolor, linkcolor=linkcolor,citecolor=citecolor]{hyperref}
\usepackage{longtable}
\usepackage{listings}

\newcommand{\JC}[1]{\color{purple}{} \color{black}}
\newcommand{\MR}[1]{\color{blue}{} \color{black}}
\newcommand{\SB}[1]{\color{olive}{} \color{black}}

\newcommand{\hn}[0]{\ensuremath{\hat n}}
\renewcommand{\a}[0]{\ensuremath{\alpha}} 
\newcommand{\va}[0]{\ensuremath{\boldsymbol{\a}}}

\newcommand{\T}[0]{\mathcal T} 
\newcommand{\D}[0]{\mathcal D} 
\newcommand{\lat}[0]{\ensuremath{\theta}}
\newcommand{\longi}[0]{\ensuremath{\varphi}}
\newcommand{\bigO}[1]{\ensuremath{\mathcal{O}(#1)}}

\newcommand{\anorm}[0]{\ensuremath{\a}} 
\newcommand{\atht}[0]{\ensuremath{\a_{\lat}}} 
\newcommand{\aphi}[0]{\ensuremath{\a_{\longi}}} 
 
\newcommand{\lmax}{\ensuremath{\ell_\text{max}}}

\newcommand{\etht}[0]{\ensuremath{\boldsymbol{e_{\lat}}}}
\newcommand{\elong}[0]{\ensuremath{\boldsymbol{e_{\longi}}}}

\newcommand{\lenspix}[0]{\texttt{LensPix}\xspace}
\newcommand{\flints}[0]{\texttt{Flints}\xspace}
\newcommand{\lenspyx}[0]{\texttt{lenspyx}\xspace}
\newcommand{\pixell}[0]{\texttt{pixell}\xspace}

\newcommand{\acceff}[0]{\ensuremath{\epsilon_{\text{eff}}}\xspace}
\newcommand{\acctarget}[0]{\ensuremath{\epsilon_{\text{target}}}\xspace}

\newcommand{\Nside}[0]{\ensuremath{N_\text{side}}}

\newcommand{\alen}[0]{\ensuremath{a^{\text{len}}_{\ell m}}\xspace}
\newcommand{\aunl}[0]{\ensuremath{a^{\text{unl}}_{\ell m}}\xspace}
\newcommand{\lmaxlen}[0]{\ensuremath{\lmax^{\text{len}}}\xspace}
\newcommand{\lmaxunl}[0]{\ensuremath{\lmax^{\text{unl}}}\xspace}

\makeatletter
\renewcommand\@makefnmark{\@textsuperscript{\normalfont\color{magenta}\@thefnmark}}
\makeatother

\usepackage{amstext}

\begin{document}

\title{Improved cosmic microwave background (de-)lensing using general spherical harmonic transforms}
\titlerunning{General SHTs for CMB lensing}
\author{Martin Reinecke\inst{\ref{mpa}}
\and Sebastian Belkner\inst{\ref{geneve}}
\and Julien Carron\inst{\ref{geneve}}}
\institute{Max-Planck Institut für Astrophysik, Karl-Schwarzschild-Str. 1, 85748 Garching, Germany\label{mpa}\and Universit\'e de Gen\`eve, D\'epartement de Physique Th\'eorique et CAP, 24 Quai Ansermet, CH-1211 Gen\`eve 4, Switzerland\label{geneve}}
\abstract{Deep cosmic microwave background polarization experiments allow a very precise internal reconstruction of the gravitational lensing signal in pricinple.
For this aim, likelihood-based or Bayesian methods are typically necessary, where very large numbers of lensing and delensing remappings on the sphere are sometimes required before satisfactory convergence.
We discuss here an optimized piece of numerical code in some detail that is able to efficiently perform both the lensing operation and its adjoint (closely related to delensing) to arbitrary accuracy, using nonuniform fast Fourier transform technology.
Where applicable, we find that the code outperforms current widespread software by a very wide margin. It is able to produce high-resolution maps that are accurate enough for next-generation cosmic microwave background experiments on the timescale of seconds on a modern laptop.
The adjoint operation performs similarly well and removes the need for the computation of inverse deflection fields.
This publicly available code enables de facto efficient spherical harmonic transforms on completely arbitrary grids, and it might be applied in other areas as well.
}


   \maketitle

\section{Motivation}
Weak gravitational lensing by large-scale structure affects radiation from the cosmic microwave background (CMB) in subtle but important ways \citep{Lewis:2006fu} by distorting and smoothing the primordial near isotropic two-point statistics and introducing a large trispectrum that can now easily be detected with very high significance \citep{Aghanim:2018oex, Carron:2022eyg, ACT:2023dou}.
Extraction of the signal is now an important piece of the science case of many CMB experiments \citep{Abazajian:2016yjj, CORE:2017ywq,SimonsObservatory:2018koc} because the lensing potential power spectrum, which probes the formation of structure at high redshift, is thought to be a particularly clean probe of the neutrino mass scale \citep{Lesgourgues:2006nd, Hall:2012kg, Allison:2015qca}.
Another reason is that removal of the lensing signal (delensing) from the CMB polarization $B$-mode \citep{Zaldarriaga:1998ar, Knox:2002pe, Kesden:2002ku} is now compulsory in order to place the best constraints on a primordial background of gravitational waves from inflation \citep{BICEP:2021xfz, Tristram:2021tvh}.

It has long been known that deep high-resolution observations of the CMB polarization in principle allow an extremely good internal reconstruction of the lensing signal \citep{Hirata:2003ka}. Recent years have seen works trying to achieve this goal of capturing a signal-to-noise ratio that is as high as possible in realistic experimental configurations \citep{Carron:2017mqf, Millea:2017fyd, Millea:2020iuw, Millea:2021had, Legrand:2021qdu, Legrand:2023jne, Aurlien:2022tlp}, and some attempts were made on data as well~\citep{POLARBEAR:2019snn, Millea:2020iuw}. These methods have in common that they use a likelihood model of the lensing signal, which allowed them to outperform the now standard quadratic estimators \citep{Hu:2001kj, Okamoto:2003zw, Maniyar:2021msb}, which are limited by the amount of lensing $B$-mode power in the data.

These likelihood-based methods form the main motivation for this work.
They are typically much more expensive than current quadratic estimator analysis: Properly modeling the subtle, $\sim 2$~arcmin remapping effect of gravitational lensing or delensing on CMB maps requires working at high resolution, and these operations must be performed many times before convergence. Numerically speaking, harmonic transforms of data distributed on the sphere are significantly costlier than in flat space. Only a handful of works were so far able to run such optimized reconstructions on large sky fractions and take the sky curvature into account~\citep{Aurlien:2022tlp, Legrand:2021qdu, Legrand:2023jne}.

In order to be slightly more concrete, we consider for example the problem of recovering the unlensed CMB from data and an estimate of the lensing deflection. The unlensed CMB likelihood conditioned on the lensing map is Gaussian, but with a covariance that is anisotropic owing to the deflection.
If no further approximation is made, recovery of the delensed CMB will involve the inverse covariance (Wiener-filtering), which today can only be performed with iterative methods such as a conjugate gradient, and needs approximately ten iterations to converge. 
Each iteration requires two remapping operations (one forward operation, and the second operation is the adjoint operation). The construction of optimal lensing mass maps internally from the CMB proceeds by iteratively applying these delensing steps and measuring residual lensing \citep{Hirata:2002jy, Hirata:2003ka}, and at best, it requires about approximately ten delensing iterations.
Hence (again, in the absence of approximations), it seems difficult to reconstruct a single best-lensing mass map with fewer than approximately 200 remapping operations. Sampling methods~\citep{Millea:2017fyd, Millea:2020iuw} are in principle orders of magnitude costlier still.

For these reasons, we have developed an optimized piece of code that is able to perform the deflection operation and its adjoint (the operations are properly defined below) efficiently.
While several pieces of software are publicly available to perform the forward-deflection operation, we find that our implementation outperforms them by far, and the adjoint operation (equally as important for likelihood-based methods) appears to be new.
\begin{figure}
   \centering
     \includegraphics[width=0.9\hsize]{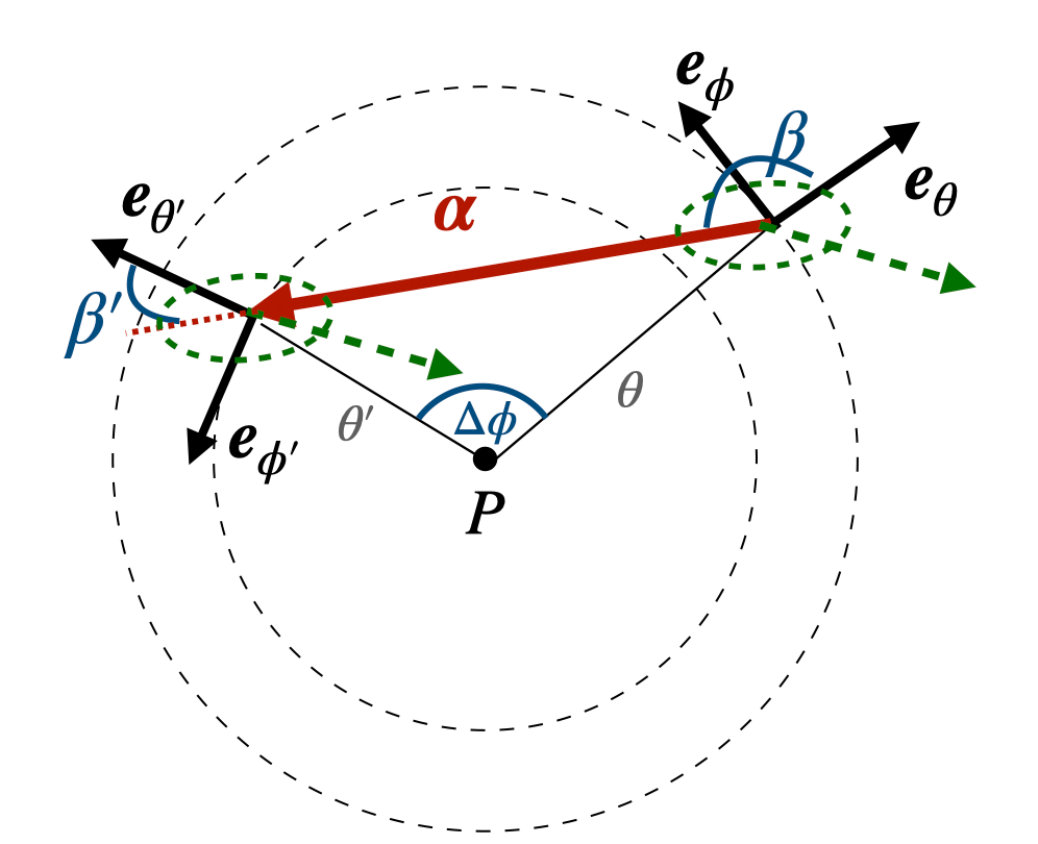}
      \caption{Lensing geometry and notation near the north pole. The sky curvature is suppressed for clarity. The deflection vector $\va(\hn)$ lies in the plane tangent to the observed coordinate $\hn$ at latitude $\lat$ and longitude $\longi$, and points toward the unlensed coordinate $\hn'$, lying a distance $\anorm = |\va(\hn)|$ away along the great circle generated by $\va$. The lensing remapping for parallel-transported spin-weighted fields like the dashed green vector or ellipse receives a phase correction $e^{i s(\beta - \beta')}$ from the rotation of the local $\lat$ and $\longi$ basis axes.}
         \label{fig:geom}
 \end{figure}
\section{Lensing and delensing the cosmic microwave background}

To a very good approximation that is valid for next-generation CMB experiments such as CMB-S4\footnote{\url{www.cmb-s4.org}} \citep{Abazajian:2016yjj}, the effect of gravitational lensing is that of a remapping of points on the sphere.
The observed CMB intensity signal at position $\hn'$ is related to that of an unlensed position by the relation
\begin{equation}
        T^\text{len}(\hn) = T^\text{unl}(\hn')
,\end{equation}
where $\hn'$ is located at a distance $\a$ from $\hn$ along the great circle generated by the deflection field $\va$.
Fig.~\ref{fig:geom} shows the geometry and our notation.
In polarization, and more generally, for any spin-weighted field $_s\T$, there is an additional phase factor that is sourced by the change in the local basis between the deflected and undeflected position (hence mostly relevant only near the poles) \citep{Challinor:2002cd}.
This may be written
\begin{equation}
\begin{split}
        _s \T^\text{len}(\hn) &= e^{i s \chi(\hn)}\:_s \T^\text{unl}(\hn') \\
\end{split}
,\end{equation}
where $\chi = \beta - \beta'$ on Fig.~\ref{fig:geom}.
We are primarily interested in efficient implementations of both the deflection operator, $\D_{\va}$,  which from a band-limited set of harmonic modes results in the lensed map on some arbitrary locations, or pixels, $\hn_i$, and of its adjoint $\D^\dagger_{\va}$.
The forward operation can be written explicitly (for spin-0 fields) as
\begin{equation}\label{eq:forward}
\left[\D_{\va} T^\text{unl} \right]_i \equiv \sum_{\ell = 0}^{\ell_\text{max}} \sum_{m = -\ell}^{\ell} T^\text{unl}_{\ell m}\:  Y_{\ell m}(\hn_i') \quad\text{ (spin-0)}
\end{equation}
and is thus closely related to the problem of finding an efficient forward spherical harmonic transform to an irregular grid.
The adjoint takes a map as input, together with a set of deflected coordinates, to produce harmonic coefficients as follows (again, here for a spin-0 field):
\begin{equation}\label{eq:adjoint}
\left[\D^\dagger_{\va} T^\text{len} \right]_{\ell m} \equiv \sum_i T^\text{len}(\hn_i)\:  Y^*_{\ell m}(\hn_i')\quad \text{  (spin-0)}
\end{equation}
for $|m |\leq \ell \leq \ell_\text{max}$. 
In the most general situation, the points $\hn_i$ and $\hn_i'$ are completely arbitrary, such that the code presented here forms in fact a spherical harmonic transform (SHT) pair that works on any pixelization of the sphere.

In situations like those encountered in CMB lensing, the points $\hn_i$ cover the sphere according to a reasonable sampling scheme (e.g., a Healpix \citep{healpix} or Gauss-Legendre grid), and $\hn_i'$ are the deflected coordinates given by Fig.~\ref{fig:geom}.
When quadrature weights are added to Eq.~\eqref{eq:adjoint}, the sum becomes an approximation to an integral over the observed coordinate,
\begin{equation}
        \left[\D^\dagger_{\va} T^\text{len} \right]_{\ell m}\sim\int d^2n \:T^\text{len}(\hn)\:  Y^*_{\ell m}(\hn')\quad \text{  (spin-0)}.
\end{equation}
This is different to the operation inverse to Eq.~\eqref{eq:forward} (delensing), as we discuss now.

If the remapping of the sphere is invertible (which is always the case in the weak-lensing regime), we can perform a variable transformation to the unlensed coordinate $\hn'$ and obtain
\begin{equation}\label{eq:adjointinverse}
\left[\D^\dagger_{\va} T^\text{len} \right]_{\ell m}\sim \int d^2n' \left(\frac{T^\text{len}(\hn)} {|A(\hn)|}\right)\:Y^*_{\ell m}(\hn')\quad \text{  (spin-0)},
\end{equation}
where $|A(\hn)| = |d^2n' / d^2n|$ is the Jacobian (magnification) determinant of the lensing remapping.

Eq.~\eqref{eq:adjointinverse} now has the form of a standard SHT of $(T / |A|)(\hn)$, where $\hn$ is matched to $\hn'$ as in Fig.~\ref{fig:geom}~(hence, $T^\text{len} / A$ is first delensed, and then mapped back to harmonic space). The choice of an isolatitude grid for $\hn'$ provides one way to calculate this integral quickly with a standard backward SHT
~(\cite{Aurlien:2022tlp, Legrand:2021qdu, Legrand:2023jne} ) implemented $\D^\dagger$ in this way.
However, significant overhead can remain with this method because it requires calculating the inverse deflection angles $\hn(\hn')$ on this grid. In a standard situation, the angles $\hn'(\hn)$ are easily obtained from a standard SHT of the deflection field on an isolatitude grid sampling the observed coordinate $\hn$, which does not provide $\hn(\hn')$ when the unlensed coordinate $\hn'$ itself is sampled on such a grid.
Moreover, usage of Eq.~\eqref{eq:adjointinverse} requires the additional calculation of the magnification determinant, which has the cost of several forward SHTs (see Appendix~\ref{app:adj_vs_inv}).
The algorithm presented here bypasses this additional work and drops the requirement of an invertible deflection field. 

For spin-weighted fields, the situation is almost identical. The harmonic space coefficients are split into a gradient ($G_{\ell m}$) and a curl term ($C_{\ell m}$), and the deflection operation is defined through
\begin{equation}\label{eq:forwardpol}
        _{s}\left[\D_{\va}\: \T \right]_i = -e^{i s \chi(\hn_i)} \sum_{\ell = 0}^{\ell_\text{max}}\sum_{m=-\ell}^\ell  \left( G_{\ell m} + i C_{\ell m} \right)\: _{s}Y_{\ell m}(\hn_i')
\end{equation}
for s > 0.
The sign is not consistent with the spin-0 case to accommodate for the most prevalent conventions in the community.
This creates a complex map of spin weight $s$, whose complex conjugate can be referred to with the subscript $-s$.

The adjoint takes this complex map as input and calculates the two sets of coefficients
\begin{equation}
_{\pm s}\left[\D^\dagger_{\va}\T \right]_{\ell m} = \sum_i e^{\mp i s\chi(\hn_i) }\: _{\pm s}\T(\hn_i) \:_{\pm s}Y^*_{\ell m}(\hn'_i)
,\end{equation}
which are decomposed as usual into gradient and curl modes,
\begin{equation}
\begin{split}
-\frac 12 &\left(\:_{s}\left[\D^\dagger_{\va}\T \right]_{\ell m} + (-1)^s \:_{-s}\left[\D^\dagger_{\va}\T \right]_{\ell m}\right) \quad (G_{\ell m})\\
-\frac 1{2i} &\left(\:_{s}\left[\D^\dagger_{\va}\T \right]_{\ell m} - (-1)^s \:_{-s}\left[\D^\dagger_{\va}\T \right]_{\ell m}\right)\quad (C_{\ell m})\end{split}
.\end{equation}
The phase $\chi$ is quite specific to CMB lensing applications and is absent in the general-purpose interpolation routines.
The relation between the adjoint and inverse is unchanged from the case of spin-0 fields.

\section{Implementation}
\label{sect:implementation}
Of the many implementations of the forward operation that were tested over the years, our approach is closest to that of \cite{Basak:2008pq}.
The fundamentals are quite straightforward: The key point is that on the sphere parameterized by co-latitude $\lat$ and longitude $\longi$, a band-limited function can be exactly written as a two-dimensional discrete Fourier series in $\lat$ and $\longi$.
While any function on the sphere is  naturally periodic in the longitude coordinate, we must artificially extend the $\theta$ range to $[0,2\pi)$ to obtain a doubly periodic function.
We can then apply nonuniform fast Fourier transform (NUFFT) techniques \citep{Barnett_2019} to this function to perform the desired interpolation.

The main steps of the forward operations can be summarized as follows:
\begin{description}
        \item [\emph{synthesis}] Synthesis of the map from $T^\text{unl}_{\ell m}$ on a rectangular equidistant grid with $n_\lat \ge \ell_\text{max}+2$  and $n_\longi \ge 2  \ell_\text{max} + 2$ points with a standard SHT, with one isolatitude ring on each pole. Both dimensions are chosen in a way to make FFTs of lengths $n_\varphi$ and
    $2n_\theta-2$ efficient; in addition, $n_\varphi$ must be an even number to enable the subsequent doubling step.
        For a standard CMB lensing application requiring $\ell_\text{max} \sim 5000$,  this corresponds to a sampling resolution close to 2 arcmin.
        The asymptotic complexity is \bigO{\ell_\text{max}^3}.
        \item [\emph{doubling}] Doubling the $\lat$-range from $[0, \pi]$ to $[0,2\pi)$, by creating a $(2n_\lat - 2, n_\longi)$ map, with the original map in the first half, and mirrored image $\lat \rightarrow 2\pi - \lat$ and $\longi \rightarrow  \longi + \pi$ in the second half.
        In the case of odd spin values, the mirrored image also takes a minus sign.
        \item [\emph{FFT}] Going to Fourier space with the help of a standard forward 2D FFT.
        These coefficients by construction contain all information necessary to evaluate the map perfectly at any location.
        The asymptotic complexity is \bigO{\ell_\text{max}^2 \log \ell_\text{max}}.
        \item [\emph{NUFFT}] Finally, from these Fourier coefficients and the coordinates $\hn'$, interpolation proper, performed with a uniform-to-nonuniform (or type 2) NUFFT.
        The asymptotic complexity is $\bigO{\ell_\text{max}^2 \log \ell_\text{max}} + \bigO{n_\text{points}}$, where $n_\text{points}$ is the number of points on the irregular grid.
        \\This is the only step that incurs inaccuracies beyond those introduced by the finite precision of floating-point arithmetics. These inaccuracies are controlled by a user-specified parameter $\epsilon$ that was described in some detail in \cite{arras-etal-2021}.
 \end{description}
The first three steps given here construct the 2D Fourier coefficients of the doubled-sphere representation of the same map from the spherical harmonic coefficients.
Our approach to performing this (\emph{synthesis-doubling-FFT}) starts with an SHT (\emph{synthesis}).
Consistent with the given asymptotic complexities, we find that this typically dominates the execution time overall.
  There are alternative approaches, however: By manipulating Wigner small-$d$ matrices, we can build a more explicit representation of the relation between spherical and 2D Fourier harmonics that can be implemented via a well-behaved three-term recurrence formula without any Fourier transforms and with a similar $\bigO{\ell_\text{max}^3}$ theoretical complexity.
  This is how \cite{Huffenberger:2010hh} implemented their SHTs, for example, and how \cite{Basak:2008pq} implemented their CMB remapping. 
  In place of a Legendre transform for each $\theta$ coordinate, a Legendre transform is required only at the equator, but one spin-weighted transform is required for each spin between 0 and $\ell_{\text{ max}}$.
  In our case, previous measurements (see, e.g., Section 2 of \citealt {beyondplanck-8-2022}) showed that it was difficult to bring this recursion to speeds on CPUs that were comparable to the highly optimized standard Legendre transforms derived from the \texttt{libsharp} library \citep{reinecke-seljebotn-2013} we are using.
  While we cannot exclude that there is some room for improvements provided the recursion can be optimized in a similar fashion (or, possibly, on GPUs), these are likely to be minor.

 For the adjoint operation, the steps naturally go backwards (the individual complexities stay unchanged):
 \begin{description}
        \item [\emph{NUFFT}] From the input map and deflected coordinates, we perform a nonuniform-to-uniform (or type 1) NUFFT, resulting in the 2D FFT Fourier coefficients.
        \item [\emph{FFT}] We remap them to position space using a standard backward FFT on the same doubled Fourier sphere of shape $(2 n_\lat-2, n_\longi)$ as for the forward operation.
        \item [\emph{undoubling}] The doubling of the Fourier sphere is undone by adding its mirror image to (or, for odd spins, subtracting from) the part $[0, \pi]$ .
        \item [\emph{adjoint synthesis}] We perform a standard adjoint SHT on this new map  of shape $(n_\lat, n_\longi)$.
        This gives us the desired spherical harmonic coefficients.
 \end{description}
Most of these steps are well established\footnote{Fast Fourier transforms are handled with code derived from the \texttt{pocketFFT} library (\url{https://gitlab.mpcdf.mpg.de/mtr/pocketfft}), which in turn is a descendant of FFTPACK (\url{https://netlib.org/fftpack/}), which was heavily modified for an improved accuracy and performance.\\
The standard ring-based spherical harmonic transforms are derived from the \texttt{libsharp} \citep{reinecke-seljebotn-2013} library (\url{https://gitlab.mpcdf.mpg.de/mtr/libsharp}).} in the astrophysical community or can be understood intuitively, with the possible exception of the NUFFT, whose purpose and structure we therefore outline 
(for more theoretical and technical details, see \cite{potts-steidl-tasche-2001,greengard-lee-2004,Barnett_2019}).
When given the Fourier coefficients of an $n$-dimensional function, it is trivial to obtain the function values on a regular grid in real space, in (almost) linear time.
This is achieved using the fast Fourier transform, potentially after zero-padding the Fourier coefficients to increase the grid resolution.
If the function values are required at irregularly spaced positions, however, this is not possible, and naive calculation of the Fourier series at each point is typically prohibitively slow in practice. 
One mathematically correct approach is to perform an FFT to a regular grid and then convolve these points with a $\mathrm{sinc}$ kernel centered on the desired locations ($\mathrm{sinc}$-interpolation).
This is just as prohibitive, but approximate solutions can be found by choosing an alternative and more suitable convolution kernel. 
In particular, a kernel with compact support will be chosen (in our case, it takes the form of eq.~(29) of \citealt{arras-etal-2021}), and
we divide the Fourier coefficients of the function by the Fourier coefficients of the kernel. This is the deconvolution step, and is followed by zero-padding the Fourier coefficients which increases the size of each dimension by a factor of roughly 1.2 to 2, and perfrom an FFT of the resulting array. At last comes the convolution step: for each irregularly spaced point, we perform a sum over its neighborhood and weight it by the kernel function.


Depending on the chosen kernel shape, zero-padding factor, and kernel support, it is possible to achieve accuracies close to machine precision;
tuning the algorithm for higher tolerances is also possible and will improve the run time considerably. 
Good kernels compromise between the conflicting properties of being fast to evaluate and having small support and a quickly decaying Fourier transform. We used the variations of the kernel proposed in \cite{Barnett_2019}, as discussed in \cite{arras-etal-2021}.

Because all steps mentioned above are linear operations, the adjoint of the described NUFFT is obtained by executing the adjoint steps in reverse order, which is the nonuniform-to-uniform, or type 1, NUFFT.

Finally, for CMB-lensing applications, the unlensed angles $\hn'$ can easily be gained from the representation
\begin{equation}
        \hn' = \cos \anorm \:\hn +  \frac{\sin \anorm}{\anorm} \left(\atht \etht + \aphi \elong \right)
,\end{equation}
where at each point, the deflection vector is obtained from the spin-1 transform  of the gradient ($\phi$) and curl ($\Omega$, if present) lensing potentials,
\begin{equation}\label{eq:deflect}
\begin{split}
        &\atht(\hn) + i\aphi(\hn) \\=- &\sum_{LM}\sqrt{L(L + 1)} \left( \phi_{LM} +  i \Omega_{LM}\right) \:_{1}Y_{L M}(\hn).
        \end{split}
\end{equation}
Here, we follow the convention of using capital letters $L M$ to refer to the spherical harmonic coefficients of the lensing potential.

\section{Benchmark}
\begin{figure}
   \centering
     \includegraphics[width=0.95\hsize]{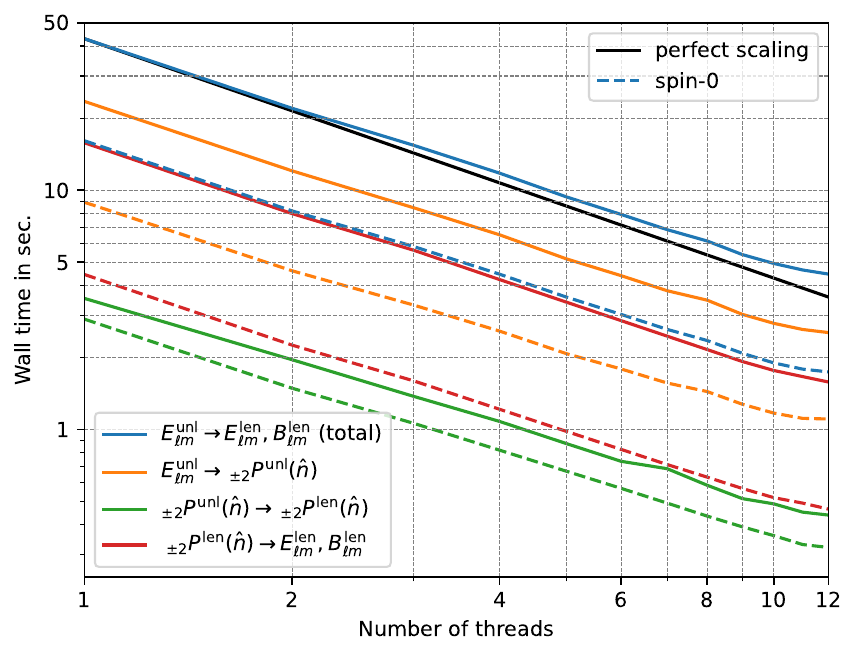}
     \includegraphics[width=0.95\hsize]{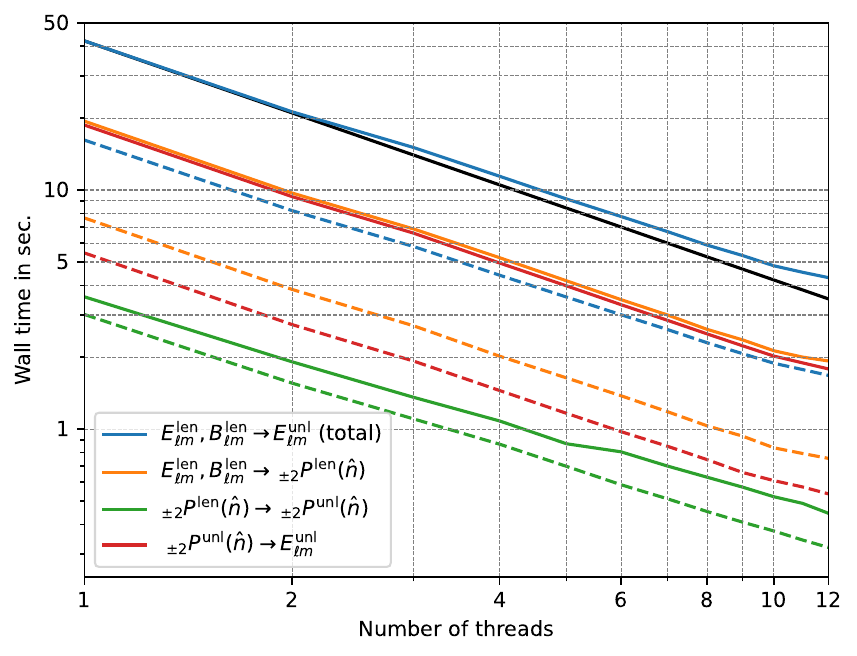}
      \caption{
      Execution time of different transforms as a function of number of threads used for the calculation.
      \emph{Upper panel:}
      Scaling of a single forward-lensing execution time, producing lensed spherical harmonic coefficients from their unlensed counterparts.
      The solid blue line shows the total time of producing $\alen$ from $\aunl$ in the polarized spin-2 case, and the solid orange, green, and red curves show the synthesis, interpolation proper and final backward SHT contributions. 
      The black line shows a perfect scaling with the inverse thread number for comparison and the scaling of each and every operation is almost perfect.
      Dashed lines show the spin-0 results, for which harmonic transforms are substantially faster.
      The forward-lensing operation is built out of remapping on a pixelized sphere (orange and green, the latter being the remapping step, properly speaking), and sending it back to harmonic space with a standard spin-2 spherical harmonic transform (red). See the text for the precise specifications.
      These curves do not include the cost of calculating the deflected angles from the deflection field spherical harmonic coefficients, which is comparable to that of a spin-1 forward spherical harmonic transform.
      \emph{Lower panel:} Corresponding results for the adjoint operation (note that while the adjoint operation is closely related to delensing, this is not the operation inverse to forward lensing; see text).
       The dashed lines show the corresponding results for the spin-0 case.}
         \label{fig:sscaling}
 \end{figure}

This section discusses the computational cost, the scaling with threading, and memory usage of the forward and adjoint operation,
and it compares our work to a few publicly available implementations of the forward operation,
such as
\lenspix\footnote{\url{https://cosmologist.info/lenspix}} \citep{Lewis:2005tp},
\lenspyx\footnote{\url{https://github.com/carronj/lenspyx}} \citep{Aghanim:2018oex},
\pixell\footnote{\url{https://pixell.readthedocs.io}} \citep{2021ascl.soft02003N},
and \flints\footnote{\url{https://bitbucket.org/glavaux/flints}} \citep{Lavaux:2010ja}.

A typical task in the CMB lensing context is to compute lensed CMB spherical harmonic coefficients (\alen) starting from unlensed ones (\aunl). 
This can be achieved by first performing a forward remapping onto a suitable pixelization of the sphere, and computing the spherical harmonics coefficients by a standard backward SHT.
The adjoint of this entire operation is first built out of a forward SHT, followed by the adjoint remapping. 
Key parameters impacting the execution time are the band limit \lmaxunl of \aunl and the requested maximum multipole \lmaxlen of the lensed CMB. 
In the applications that motivated this work, \lmaxlen is at most the multipole above which the information on the lensing signal becomes negligible. 
For example, \textit{Planck}~reconstructions \citep{Aghanim:2018oex, Carron:2022eyg} used $\ell^{\rm len}_{\text{ max}} = 2048$, and the recent ACT results~\citep{ACT:2023kun} $\lmaxlen = 3000$. For a very deep future polarization experiment such as the CMB-S4 deep survey \citep{Abazajian:2016yjj}, this is closer to $\ell^{\rm len}_{\text{ max}} = 4096$, a number we use often as reference in this section. 
Then \lmaxunl must typically be taken slightly higher than \lmaxlen in order to account for the mode mixing by lensing.
 
When lensed spherical harmonic coefficients are built in this way, another important parameter is the intermediate grid pixelization. 
Because the lensed CMB is not band limited, there is no exact quadrature rule, and this choice can strongly impact the accuracy of the recovered \alen at high multipoles. 
We show in Fig.~\ref{fig:sscaling} the execution time of these tasks for the forward (upper panel) and adjoint (lower panel) cases as a function of the number of threads (strong scaling).
We picked   $\lmaxunl=5120$, $\lmaxlen=4096$, and a longitude-thinned Gauss-Legendre grid with 6144 rings as intermediate grid ($\sim 5\cdot 10^{7}$ pixels). This is a conservative configuration that in our experience allows a robust and accurate lensing reconstruction for very deep stage IV CMB observations. 
This was performed on a CSCS Piz Daint XC50 compute node\footnote{\url{https://www.cscs.ch/computers/piz-daint/}}, with 12 physical cores. Results on more recent processors can sometimes be up to twice as fast in our experience. 
The top panel shows the result for the synthesis operation, while the lower panels shows the adjoint of it, and we see that each operation scales almost perfectly.
We have excluded the cost of building the undeflected angles from the timing, which we briefly discuss below, as is suitable in the context of optimal lensing reconstruction, where many maps are deflected with the same set of angles. 
The accuracy of the interpolation was set to $10^{-5}$, which value is good enough for most purposes, as also discussed in more detail below. 
The interpolation is very efficient. Only a minor part of the total time is dominated by the pair of SHTs that is involved.

We now discuss the accuracy and memory usage of our implementation. Our code uses simple heuristics to allow the user to choose a target-relative accuracy, \acctarget, of at least $10^{-13}$.
For the forward operation, schematically $a_{lm}^{\rm unl} \rightarrow \tilde{m}$,
we calculated a map-level effective relative accuracy, \acceff, as follows: We took the difference between the true and estimated lensed maps,
$\tilde{m}^{\rm true}$, $\tilde{m}^{\rm est}$, respectively,
and normalized by the total power of the true lensed maps,
\begin{align}
\acceff = \frac{1}{P^{\rm true}}\sqrt{\sum_i\left(\tilde{m}_i^{\rm true}-\tilde{m}_i^{\rm est}\right)^2}\,.
\end{align}
Here, $P^{\rm true}=\sqrt{\sum_i{(m_i^{\rm true})}^2}$, and the sum index $i$ run over pixels as well as the map components for nonzero spin ($Q$ and $U$ in polarization). The maps $\tilde{m}^{\rm true}$ were determined by a brute-force approach:
We remapped the unlensed map to the desired grid by calculating eq.~\eqref{eq:forward} explicitly.
As this can be fairly expensive, we calculated $10^5$ exactly lensed pixels at most for a few numbers of isolatitude rings close to the equator.

\begin{figure}
          \includegraphics[width=0.99\columnwidth]{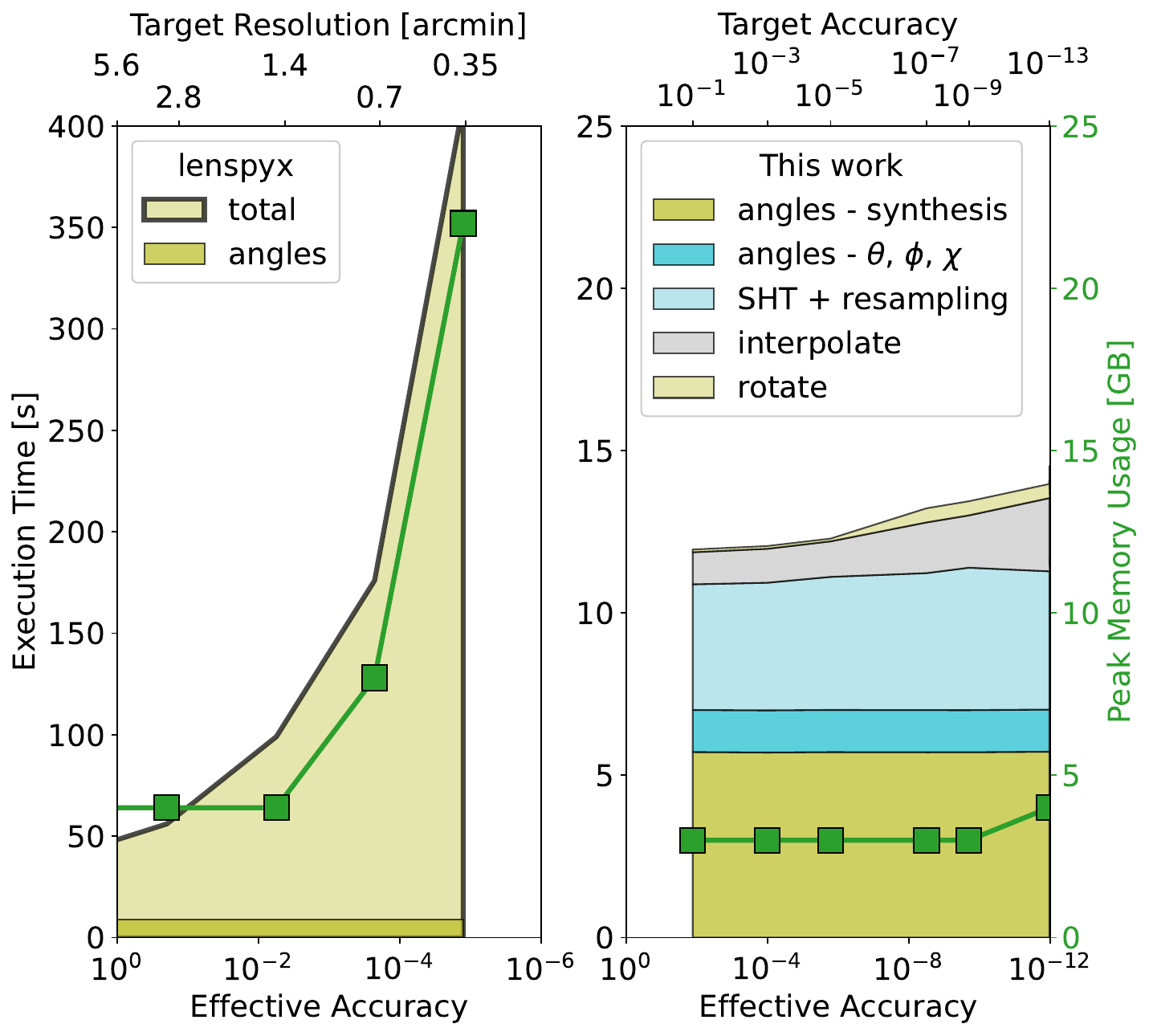}
           \caption{
                Full-sky forward operation execution times for spin-$2$ fields of our implementation compared with \lenspyx, which uses a popular bicubic spline interpolation technique.  
                The top panels show \lenspyx (left) and the new implementation (right) as a function of the effective relative accuracy for $\ell_{\text{max}} = 4096$, producing lensed CMB maps in a Healpix geometry with $\Nside=2048$.
                The effective accuracy is calculated for a few isolatitude rings close to the equator.
                All calculations were made with four CPUs.
                We indicate the execution time (shaded areas, left y-axes), peak memory usage (green data points, right y-axis),
                and the accuracy parameters (target resolution, target accuracy) at the top x-axes.
                For \lenspyx, we divided the full sky map into two bands to reduce the memory consumption.
                The differently colored areas show the main steps of the operation. The execution time and memory consumption both grows rapidly for \lenspyx\ because it requires high-resolution grids for an accurate interpolation.
           }\label{fig:bmlenspyx}
  \end{figure}

The right panel in Figure \ref{fig:bmlenspyx} shows the computational cost of the lensing routine for spin-$2$ fields at $\lmaxunl = 4096$, using four CPUs, mapping onto a Healpix pixelization with $\Nside=2048$, as a function of \acceff (bottom axis) and \acctarget (top axis). 
The effective accuracy is typically slightly higher than requested, except in the vicinity of $10^{-13}$. 
This is not a problem of the NUFFT interpolation, however, but rather of the true SHT calculations themselves, which lose several digits in accuracy at this band limit.
The diagram shows the split of the total computational cost into its most relevant steps described earlier. 
The SHT steps for the angle and doubled Fourier sphere always dominate the cost in this configuration, and the choice of accuracy has a fairly minor impact.
If relevant, the SHTs can be performed with a gradient-only setting for nonzero spin fields (an implementation specific to the case of a vanishing curl component), further reducing the computational cost of these tasks by about 25\%.
The scale of the peak memory usage is indicated on the right y-axis of the right panel, and is shown as connected green square data points.
The timings, memory, and accuracy calculations were made for a set of five analyses, and their standard deviation is negligible.

The left panel shows the corresponding results for a code using a popular interpolation scheme, \lenspyx. 
This code (just as \lenspix and \pixell) uses bicubic spline interpolation on a intermediate grid obtained by cylindrical projection.
The resolution of this grid then essentially sets the accuracy of the result and memory usage. 
While these interpolation schemes are perfectly fine for the purpose of producing a set of lensed CMB maps, they have strong limitations for more intensive tasks, or when higher accuracies are imperative.
For low target resolutions of about $1.4$ arcmin, the new implementation speeds up the execution time approximately 7 times for a similarly effective accuracy of about $\acceff\approx 10^{-2}$.
This increases to a speed-up of 30 times for an effective accuracy of $10^{-5}$.
Higher accuracies are almost not manageable for \lenspyx in both time and memory, whereas the new implementation can easily reach accuracies as low as $10^{-12}$.

\begin{figure}
                \includegraphics[width=0.99\columnwidth]{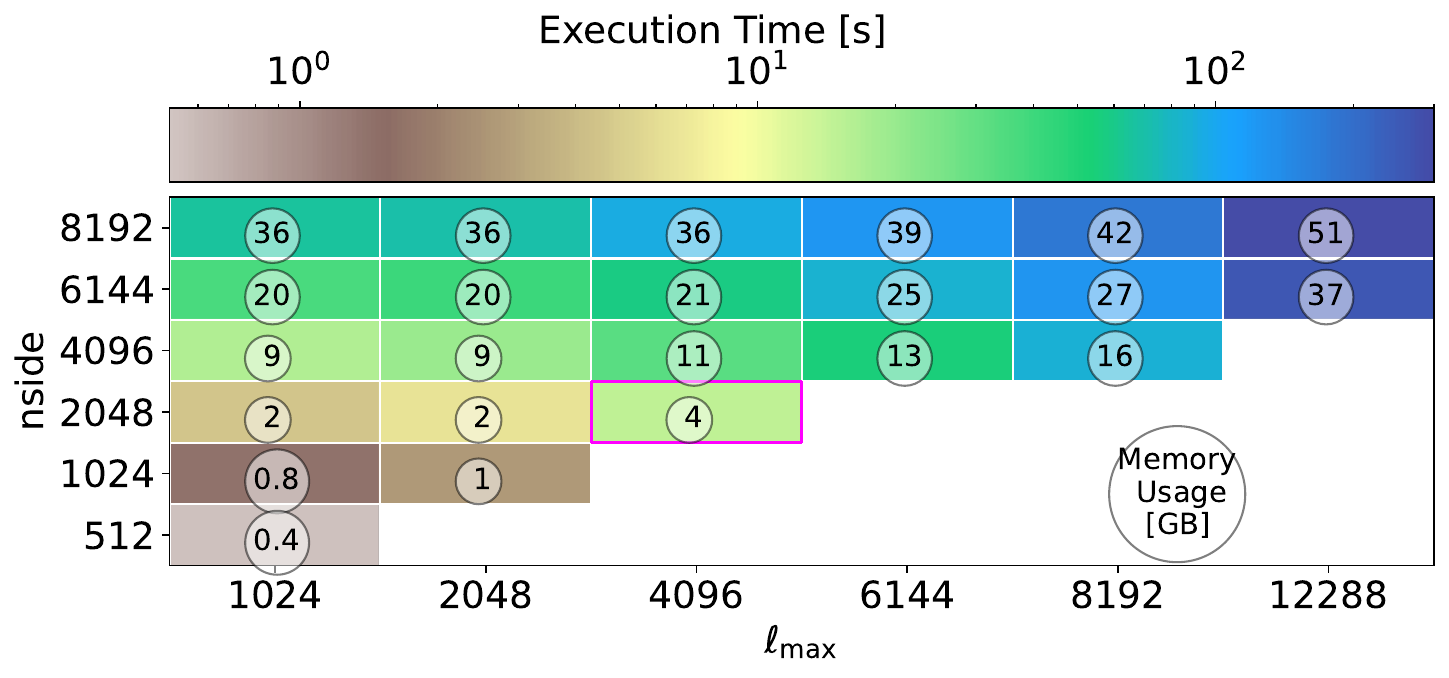}
                \caption{
                Benchmark of the execution time and peak memory consumption for a wide range of Healpix resolution \Nside~and $\ell_{\text{max}}$, for a target accuracy of $10^{-13}$, and for our new implementation. The cell highlighted in magenta is the configuration we used for the accuracy benchmark in Fig.~\ref{fig:bmlenspyx}.
                The execution time ranges from $0.55$ seconds ($\Nside = 512, \ell_{\text{max}}=1024$) to $336$ seconds ($\Nside = 8192, \ell_{\text{max}} = 12288$), while the memory consumption increases from $0.4$ GB to $51$ GB. The memory consumption closely follows the memory needed to store the Healpix maps of that particular size.
                }\label{fig:bmlenspyxscan}
        \end{figure}

Figure \ref{fig:bmlenspyxscan} shows the execution time and peak memory usage of the new implementation for various \Nside\ and $\lmaxunl$ configurations and a target accuracy of $\acctarget=10^{-13}$.
The memory usage for the most challenging benchmarks is still below $64$ GB, and we would like to note that the memory consumption is only slightly larger than the memory needed for the three maps (the deflection angles) for the respective \Nside.

We now compare our results to other implementations that benchmark the spin-$0$ forward operation.
The effective accuracy was calculated as in the above case, replacing the data by the spin-$0$ true and estimated temperature maps.
For \lenspix and \flints, we estimated the costs by using the natively C-implemented lensing routine, and with \texttt{OpenMP} support and four CPUs.
We measured the tasks directly related to lensing, such as the calculation of deflection angles, the deflection, the interpolation, and rotation.
To farily compare all implementations, we did not cache any calculations in \flints, even though they are available.
For \pixell, we used the full-sky lensing routine implemented in Python, and measured the costs within Python.

Almost all implementations provide an accuracy parameter in form of the intermediate grid pixel resolution. We chose them to provide approximately comparable results.

For \flints, \lenspix, and \lenspyx, we generated input CMB and lensing potential realizations and the true and estimated lensed maps on a Healpix grid.
For \pixell, the full-sky lensing routine was evaluated onto a CAR geometry, and we calculated the effective accuracy using our own CAR geometry to calculate the true lensed map.

We used $\lmaxunl=4096$ and an HPC machine with four CPUs and 60 GB of memory.
To reduce memory usage, the data for \lenspix were split into sets and were calculated individually.

\begin{table*}[htb!]
        \centering
    \caption{spin-0, $\ell_{\text{max}}=4096$, and $\Nside=2048$ (if applicable) forward-operation execution time and memory consumption for different implementations on the full sky, using 4 CPUs.
        Additionally, we split the full sky map for \pixell~and \lenspyx~into two bands to reduce memory consumption.}
    \begin{tabular}{llrrr}
    \hline
                implementation  &       grid resolution & effective accuracy            & computation time        & memory peak usage \\
                \hline                                  
                \pixell         & $1.0$ arcmin  & $6\cdot 10^{-3}$              & 3 min 30 sec    & 20 GB \\
                \lenspix        & $0.85$ arcmin & $3\cdot 10^{-4}$              & 9 min 30 sec    & 3.4 GB \\     
                \flints         & N.A.                  & $3\cdot 10^{-3}$              & 4 min                   & 6.5 GB \\
                \lenspyx        & $0.70$ arcmin & $1\cdot 10^{-4}$              & 2 min                   & 8 GB \\
                                        & $0.35$ arcmin & $1\cdot 10^{-5}$              & 5 min                   & 22 GB \\
                This work       & N.A.                  & $2\cdot 10^{-6}$              & 9 sec                   & 2.8 GB \\
                                        & N.A.                  & $4\cdot 10^{-12}$               & 10 sec                        & 2.8 GB \\
        \hline
        \end{tabular}\label{tab:fwopimplementations}
    \end{table*}

Table~\ref{tab:fwopimplementations} summarizes the computational costs and memory usage for the pixel resolution and effective accuracy.
\lenspix and \flints are well-established algorithms and perform the task in reasonable time and for very reasonable memory usage.
\lenspix allows us to control the intermediate grid pixel resolution, whereas with \flints, this can be indirectly controlled by the choice of \Nside.
If needed, \flints execution time could be reduced by caching some of the calculations, making them available for repeated runs.
\pixell supports splitting of the full sky data into bands, which can be used to reduce the memory usage, if needed.
The algorithm in this paper, shown in the bottom row, is vastly more efficient both in execution time and memory consumption,
even more so in the high-accuracy regime.

Finally, we comment more specifically on the relevance of this code for a maximum a posteriori lensing reconstruction \citep{Hirata:2003ka, Carron:2017mqf}.
In these methods, the reconstruction proceeds by finding an approximate solution to the Wiener-filtered delensed CMB using a conjugate gradient solver at each iteration that involves applying the forward and adjoint operation.
As mentioned earlier, this typically involves several hundred such operations.
The benefits for a reconstruction strategy like this are twofold.
Faster operations directly speed up the lensing reconstruction, and accurate operations can prevent the conjugate gradient solver from building up errors or showing instability.

We saw this explicitly by testing the full-sky lensing reconstruction with CMB-S4-like settings ($(\lmaxunl,\lmaxlen)=(4500,4000)$, a polarization noise level of $\sqrt{2}$ arcmin, and 4 CPUs), reconstructing the lensing potential using polarization-only maps, using both \lenspyx, and the new implementation for the forward and adjoint operations. 
With \lenspyx, we observed that a target resolution of about $1.0$ arcmin is beneficial for a stable reconstruction of the very largest modes of the lensing potential in experimental configurations like these, resulting in execution times for a single adjoint operation of about 3 minutes.
This is significantly longer than for the new implementation, for which we find execution times of only $8$ seconds for better accuracy.
Analogously, we find for the execution time of a typical full-iteration step of a lensing reconstruction $\sim 25$ minutes, or $\sim 4$ minutes, respectively,
and the speed-up is even larger for lower $\lmax$ analyses ($\sim 24$ minutes vs. $\sim 2$ for $((\lmaxunl,\lmaxlen)=(3200,3000)$).
It is worth mentioning that with \lenspyx at this resolution, outliers can occasionally occur. They can take more than $\text{one}$ hour or do not converge at all. This is directly due to the lower accuracy, in particular very close to the poles, where the bicubic spline method is less accurate. 
This has not been observed with the new implementation, whose error map is uniform across the entire sphere.  
We observed something similar for temperature-only reconstruction, where an effective relative accuracy of $10^{-7}$  generally appears to be required for a successful convergence of the lensing-map search on the very largest scales. This accuracy is accessible only with the implementation discussed here.

\section{Conclusion}
We have described an optimized implementation of the spherical transform pair of an arbitrary spin-weight that can be used on any pixelization of the sphere, such as regular grids distorted by CMB lensing.
A C++ implementation and comprehensive Python front-end is available, together with the low-level algorithms (FFT, NUFFT, and SHT), under the terms of the GNU General Public License and named \texttt{DUCC}\footnote{\url{https://gitlab.mpcdf.mpg.de/mtr/ducc}}.
The code is written with the goal of portability and does not depend on external libraries.
It supports multithreading via the C++ standard threading functionality and will make use of CPU vector instructions on \texttt{x86} and \texttt{ARM} hardware, if the compiler supports the respective language extensions.
The Python interface is kept deliberately general and flexible to allow use in the widest possible range of scientific applications. As a consequence, parts of the interface are somewhat complex and are perhaps best used by higher-level more application-specific packages to hide unnecessary details from the end user.

For users interested in applications specific to CMB lensing, the \lenspyx\footnote{\url{https://github.com/carronj/lenspyx}} Python package has been updated in this spirit to include these developments and provide additional wrappers to these routines. This results in an improvent of some orders of magnitude in execution time and accuracy over currently publicly available tools.

\begin{acknowledgements}
We thank Guilhem Lavaux, Antony Lewis, and Mathew Madhavacheril for discussions on \flints, \lenspix and \pixell respectively.
JC and SB acknowledge support from a SNSF Eccellenza Professorial Fellowship (No. 186879).
This work was supported by a grant from the Swiss National Supercomputing Centre (CSCS) under project ID s1203.
\end{acknowledgements}

\bibliographystyle{aa}
\bibliography{julbib,mrbib}

\begin{thebibliography}{43}
\expandafter\ifx\csname natexlab\endcsname\relax\def\natexlab#1{#1}\fi

\bibitem[{Abazajian {et~al.}(2016)}]{Abazajian:2016yjj}
Abazajian, K.~N. {et~al.} 2016 [\eprint[arXiv]{1610.02743}]

\bibitem[{Adachi {et~al.}(2020)}]{POLARBEAR:2019snn}
Adachi, S. {et~al.} 2020, Phys. Rev. Lett., 124, 131301

\bibitem[{Ade {et~al.}(2019)}]{SimonsObservatory:2018koc}
Ade, P. {et~al.} 2019, JCAP, 02, 056

\bibitem[{Ade {et~al.}(2021)}]{BICEP:2021xfz}
Ade, P. A.~R. {et~al.} 2021, Phys. Rev. Lett., 127, 151301

\bibitem[{Aghanim {et~al.}(2020)}]{Aghanim:2018oex}
Aghanim, N. {et~al.} 2020, Astron. Astrophys., 641, A8

\bibitem[{Allison {et~al.}(2015)Allison, Caucal, Calabrese, Dunkley, \& Louis}]{Allison:2015qca}
Allison, R., Caucal, P., Calabrese, E., Dunkley, J., \& Louis, T. 2015, Phys. Rev. D, 92, 123535

\bibitem[{{Arras} {et~al.}(2021){Arras}, {Reinecke}, {Westermann}, \& {En{\ss}lin}}]{arras-etal-2021}
{Arras}, P., {Reinecke}, M., {Westermann}, R., \& {En{\ss}lin}, T.~A. 2021, Astron. Astrophys., 646, A58

\bibitem[{Aurlien {et~al.}(2022)}]{Aurlien:2022tlp}
Aurlien, R. {et~al.} 2022 [\eprint[arXiv]{2211.14342}]

\bibitem[{Barnett {et~al.}(2019)Barnett, Magland, \& af~Klinteberg}]{Barnett_2019}
Barnett, A.~H., Magland, J., \& af~Klinteberg, L. 2019, {SIAM} Journal on Scientific Computing, 41, C479

\bibitem[{Basak {et~al.}(2008)Basak, Prunet, \& Benabed}]{Basak:2008pq}
Basak, S., Prunet, S., \& Benabed, K. 2008, in {12th Marcel Grossmann Meeting on General Relativity}, 2213--2215

\bibitem[{Carron \& Lewis(2017)}]{Carron:2017mqf}
Carron, J. \& Lewis, A. 2017, Phys. Rev. D, 96, 063510

\bibitem[{Carron {et~al.}(2022)Carron, Mirmelstein, \& Lewis}]{Carron:2022eyg}
Carron, J., Mirmelstein, M., \& Lewis, A. 2022, JCAP, 09, 039

\bibitem[{Challinor \& Chon(2002)}]{Challinor:2002cd}
Challinor, A. \& Chon, G. 2002, Phys. Rev. D, 66, 127301

\bibitem[{Challinor {et~al.}(2018)}]{CORE:2017ywq}
Challinor, A. {et~al.} 2018, JCAP, 04, 018

\bibitem[{{Galloway} {et~al.}(2022){Galloway}, {Reinecke}, {Andersen}, {Aurlien}, {Banerji}, {Bersanelli}, {Bertocco}, {Brilenkov}, {Carbone}, {Colombo}, {Eriksen}, {Foss}, {Franceschet}, {Fuskeland}, {Galeotta}, {Gerakakis}, {Gjerl{\o}w}, {Hensley}, {Herman}, {Iacobellis}, {Ieronymaki}, {Ihle}, {Jewell}, {Karakci}, {Keih{\"a}nen}, {Keskitalo}, {Maggio}, {Maino}, {Maris}, {Paradiso}, {Partridge}, {Suur-Uski}, {Svalheim}, {Tavagnacco}, {Thommesen}, {Watts}, {Wehus}, \& {Zacchei}}]{beyondplanck-8-2022}
{Galloway}, M., {Reinecke}, M., {Andersen}, K.~J., {et~al.} 2022, arXiv e-prints, arXiv:2201.03478

\bibitem[{{G{\'o}rski} {et~al.}(2005){G{\'o}rski}, {Hivon}, {Banday}, {Wand elt}, {Hansen}, {Reinecke}, \& {Bartelmann}}]{healpix}
{G{\'o}rski}, K.~M., {Hivon}, E., {Banday}, A.~J., {et~al.} 2005, \apj, 622, 759

\bibitem[{Greengard \& Lee(2004)}]{greengard-lee-2004}
Greengard, L. \& Lee, J.-Y. 2004, SIAM Review, 46, 443

\bibitem[{Hall \& Challinor(2012)}]{Hall:2012kg}
Hall, A.~C. \& Challinor, A. 2012, Mon. Not. Roy. Astron. Soc., 425, 1170

\bibitem[{Hirata \& Seljak(2003{\natexlab{a}})}]{Hirata:2002jy}
Hirata, C.~M. \& Seljak, U. 2003{\natexlab{a}}, Phys. Rev., D67, 043001

\bibitem[{Hirata \& Seljak(2003{\natexlab{b}})}]{Hirata:2003ka}
Hirata, C.~M. \& Seljak, U. 2003{\natexlab{b}}, Phys. Rev., D68, 083002

\bibitem[{Hu \& Okamoto(2002)}]{Hu:2001kj}
Hu, W. \& Okamoto, T. 2002, Astrophys. J., 574, 566

\bibitem[{Huffenberger \& Wandelt(2010)}]{Huffenberger:2010hh}
Huffenberger, K.~M. \& Wandelt, B.~D. 2010, Astrophys. J. Suppl., 189, 255

\bibitem[{Kesden {et~al.}(2002)Kesden, Cooray, \& Kamionkowski}]{Kesden:2002ku}
Kesden, M., Cooray, A., \& Kamionkowski, M. 2002, Phys. Rev. Lett., 89, 011304

\bibitem[{Knox \& Song(2002)}]{Knox:2002pe}
Knox, L. \& Song, Y.-S. 2002, Phys. Rev. Lett., 89, 011303

\bibitem[{Lavaux \& Wandelt(2010)}]{Lavaux:2010ja}
Lavaux, G. \& Wandelt, B.~D. 2010, Astrophys. J. Suppl., 191, 32

\bibitem[{Legrand \& Carron(2022)}]{Legrand:2021qdu}
Legrand, L. \& Carron, J. 2022, Phys. Rev. D, 105, 123519

\bibitem[{Legrand \& Carron(2023)}]{Legrand:2023jne}
Legrand, L. \& Carron, J. 2023 [\eprint[arXiv]{2304.02584}]

\bibitem[{Lesgourgues \& Pastor(2006)}]{Lesgourgues:2006nd}
Lesgourgues, J. \& Pastor, S. 2006, Phys. Rept., 429, 307

\bibitem[{Lewis(2005)}]{Lewis:2005tp}
Lewis, A. 2005, Phys. Rev., D71, 083008

\bibitem[{Lewis \& Challinor(2006)}]{Lewis:2006fu}
Lewis, A. \& Challinor, A. 2006, Phys. Rept., 429, 1

\bibitem[{Lewis {et~al.}(2002)Lewis, Challinor, \& Turok}]{Lewis:2001hp}
Lewis, A., Challinor, A., \& Turok, N. 2002, Phys. Rev. D, 65, 023505

\bibitem[{Madhavacheril {et~al.}(2023)}]{ACT:2023kun}
Madhavacheril, M.~S. {et~al.} 2023 [\eprint[arXiv]{2304.05203}]

\bibitem[{Maniyar {et~al.}(2021)Maniyar, Ali-Ha\"\i{}moud, Carron, Lewis, \& Madhavacheril}]{Maniyar:2021msb}
Maniyar, A.~S., Ali-Ha\"\i{}moud, Y., Carron, J., Lewis, A., \& Madhavacheril, M.~S. 2021, Phys. Rev. D, 103, 083524

\bibitem[{Millea {et~al.}(2019)Millea, Anderes, \& Wandelt}]{Millea:2017fyd}
Millea, M., Anderes, E., \& Wandelt, B.~D. 2019, Phys. Rev. D, 100, 023509

\bibitem[{Millea \& Seljak(2022)}]{Millea:2021had}
Millea, M. \& Seljak, U. 2022, Phys. Rev. D, 105, 103531

\bibitem[{Millea {et~al.}(2021)}]{Millea:2020iuw}
Millea, M. {et~al.} 2021, Astrophys. J., 922, 259

\bibitem[{{Naess} {et~al.}(2021){Naess}, {Madhavacheril}, \& {Hasselfield}}]{2021ascl.soft02003N}
{Naess}, S., {Madhavacheril}, M., \& {Hasselfield}, M. 2021, {Pixell: Rectangular pixel map manipulation and harmonic analysis library}, Astrophysics Source Code Library, record ascl:2102.003

\bibitem[{Okamoto \& Hu(2003)}]{Okamoto:2003zw}
Okamoto, T. \& Hu, W. 2003, Phys. Rev., D67, 083002

\bibitem[{Potts {et~al.}(2001)Potts, Steidl, \& Tasche}]{potts-steidl-tasche-2001}
Potts, D., Steidl, G., \& Tasche, M. 2001, Fast Fourier Transforms for Nonequispaced Data: A Tutorial, ed. J.~J. Benedetto \& P.~J. S.~G. Ferreira (Boston, MA: Birkh{\"a}user Boston), 247--270

\bibitem[{Qu {et~al.}(2023)}]{ACT:2023dou}
Qu, F.~J. {et~al.} 2023 [\eprint[arXiv]{2304.05202}]

\bibitem[{{Reinecke} \& {Seljebotn}(2013)}]{reinecke-seljebotn-2013}
{Reinecke}, M. \& {Seljebotn}, D.~S. 2013, Astron. Astrophys., 554, A112

\bibitem[{Tristram {et~al.}(2022)}]{Tristram:2021tvh}
Tristram, M. {et~al.} 2022, Phys. Rev. D, 105, 083524

\bibitem[{Zaldarriaga \& Seljak(1998)}]{Zaldarriaga:1998ar}
Zaldarriaga, M. \& Seljak, U. 1998, Phys. Rev. D, 58, 023003

\end{thebibliography}

\begin{appendix}
        \section{Map analysis with a minimal number of rings}

In the text above, we regularly use equiangular spherical grids with $\lmax+2$ rings (first and last ring located at the poles) to represent functions with a band limit of \lmax\ (inclusive).
This number of rings may sound insufficient to fully represent a function with the given band limit because the minimum number of rings required for an accurate map analysis via a quadrature rule on this layout is $2\lmax+2$ (Clenshaw-Curtis quadrature).

However, we can again make use of the double Fourier sphere technique that was introduced in Sect.~\ref{sect:implementation}, that is,\ we can follow a meridian from the north pole to the south pole, and then back again on the opposite side. A full meridian like this has $2\lmax+2$ points in total.
Because we assumed the function on the sphere to be band limited, the $\theta$-dependent function along each of these full meridians is the sum of associated Legendre polynomials of degrees up to \lmax\ in $\cos(\theta)$.
In other words, it can be expressed as a Fourier series $\sum_{k=0}^{\lmax} A_k \cos(k\theta)$.
This function in turn is completely determined by $2\lmax+1$ equidistant samples in the range $\theta=[0; 2\pi)$, that is,\ one less than we actually have.
The same is true in azimuthal direction, where we also have at least $2\lmax+2$ pixels on each ring.

As a consequence, the actual function value at any $\theta$ and $\varphi$ can be obtained using a combination of fast Fourier transforms and phase-shifting factors, or (in an approximate fashion) via NUFFT.
One way of extracting the spherical harmonic coefficients from this map would therefore be
by first computing the function values at a shifted set of isolatitude rings located exactly between the existing ones, increasing the number of rings to $2\lmax+3$. This is then followed by applying the appropriate Clenshaw-Curtis quadrature weights to the full set of rings. Finally, running an adjoint spherical harmonic synthesis operation on the full set of weighted rings gives the desired result.

It is even possible to shift the newly generated rings back to the original positions after weighting, which again reduces the number of rings in the adjoint synthesis operation to $\lmax+2.$ This speeds up the SHT considerably.

\section{Adjoint and inverse lensing}\label{app:adj_vs_inv}
In the most general situation, invertibility of the deflection field is not necessarily always achieved at all points in a likelihood search under nonideal conditions, where lensing estimators can react strongly to signatures of anisotropies unrelated to lensing. When the lensing deflection is invertible, the inverse lensing operation can still be useful. To this end, we can use the same adjoint operation $\D^\dagger$, but with input $T^\text{len}(\hn) \cdot |A|(\hn)$ instead of $T^\text{len}(\hn)$. 
Eq.~\eqref{eq:adjointinverse} in the main text shows that the result then is $T_{\ell m}^\text{unl}$. 

The magnification determinant may be obtained as follows:
With $_1\alpha = \atht + i \aphi$ as in Eq.~\eqref{eq:deflect}, let the convergence ($\kappa$), field rotation ($\omega$) and shears ($\gamma$) be
\begin{equation}
\begin{split}
        \kappa + i \omega &= \frac 12 \tilde \eth\: _{1}\alpha \\ \gamma_1 + i\gamma_2 &= \frac 12 \eth\: _{1}\alpha,
\end{split}
\end{equation}
where $\eth$ and $\tilde \eth$ are the spin raising and lowering operators (see the first appendix of \cite{Lewis:2001hp} for a discussion in the context of the CMB).
It holds \begin{equation}\begin{split}|A|(\hn) &= \frac{\sin \alpha }{\alpha}\left((1 -\kappa)^2 + \omega^2 - \gamma^2\right) \\&+ \left(\cos \alpha - \frac{\sin \alpha }{\alpha} \right)\left( 1 - \kappa -\cos(2\beta) \gamma_1 - \sin(2\beta) \gamma_2 \right).\end{split}\end{equation}
All of these quantities can be computed from the harmonic coefficients of the deflection field with the help of spin-weighted spherical harmonic transforms. 

Informally, ignoring technical issues of band limits, quadrature weights, and so on, we may write
\begin{equation}
        \left[\D_{\va}\D^\dagger_{\va}\right](\hn_i, \hn_j) \sim \frac{\delta^{D}(\hn_i - \hn_j)}{|A(\hn_i)|}, \quad \text{($\va$ invertible)}
,\end{equation}
where $\delta^D$ is the Dirac delta. Similarly, the operator $\D^\dagger \D$ produces the spherical harmonic coefficients of $\left(T^{\text{unl}}/|A|\right)(\hn)$ from those of $T^{\text{unl}}$.

\end{appendix}

\end{document}